# The Metal-Insulator Transition in $Y_{1-x}Ca_xTiO_3$ Caused by Phase Separation

K. Kato[1], E. Nishibori[2], M. Takata[2], M. Sakata[2],

T. Nakano[3], K. Uchihira[3], M. Tsubota[3], F. Iga[3] and T. Takabatake[3]

JASRI[1], Dept. of Appl. Phys., Nagoya Univ.[2], ADSM, Hiroshima Univ.[3]

In order to explore the origin of the Metal-Insulator (M-I) transition, the precise crystal structures of the hole-doped Mott insulator system, $Y_{1-X}Ca_XTiO_3$ (x=0.37, 0.39 and 0.41), are studied for the temperature range between 20K and 300K by the synchrotron radiation (SR) X-ray powder diffraction. For both $Y_{0.63}Ca_{0.37}TiO_3$ and $Y_{0.61}Ca_{0.39}TiO_3$ compositions, the orthorhombic (*Pbnm*) – monoclinic (*P2₁/n*) structural phase transition occurs at around 230K, which is much higher than their own M-I transition temperatures, i.e. 60K and 130K, respectively. For these compositions, the significant phase separation (low-temperature orthorhombic + monoclinic) is also found in the vicinity of the M-I transition temperature. On the other hand, $Y_{0.59}Ca_{0.41}TiO_3$, which dose not exhibit M-I transition and preserve a metallic behavior down to 1.5K, is in two phases state from 20K to 300K. It is concluded that the existence of the phase separation causes the M-I transition in $Y_{1-x}Ca_xTiO_3$, and the low-temperature orthorhombic phase contributes to the metallic property of this system.

Perovskite-type Ti-oxide, $YTiO_3$, is well known as a typical Mott-Hubbard insulator with $3d^1$ (Ti) configuration. It has been reported from the electrical resistivity using polycrystalline samples that a hole-doping effect on Ti site by Ca substitution for Y causes a Metal-Insulator (M-I) transition at around x=0.4 in $Y_{1-X}Ca_XTiO_3$ [1]. It was not possible to determine the definite temperature and the doping Ca content to make the polycrystalline samples the M-I transition, because the transition observed in these samples was blurred due to the inevitable inhomogeneity of a mixed-crystal system [2]. Recently, F. Iga *et al.* succeeded in synthesizing single crystals of $Y_{1-X}Ca_XTiO_3$ (x=0.37, 0.39 and 0.41) for the first time [3]. From the electrical resistivity measured by the single crystals, the M-I transition temperatures of this system were clearly determined and found to change systematically depending on the doping Ca content [3]. So far, in order to see the origin of the M-I transition, this hole-doped Mott insulator system has been investigated extensively using not only polycrystalline samples but also single crystals by the various experimental techniques [4,5,6,7,8,9]. However, the precise crystal structure investigation relating to the M-I transition has never been done. Hence it is not possible to answer the fundamental question such as whether there exists any phase transition related to the M-I transition or not. In this study, the precise crystal structures of the hole-doped Mott insulator system, $Y_{1-X}Ca_XTiO_3$ (x=0.37, 0.39 and 0.41), are revealed for the temperature range between 20K and 300K by the high-resolution X-ray powder diffraction experiment using synchrotron radiation (SR). Through the present structural study, the conclusive experimental evidence of the M-I transition is obtained. From the evidence, it is obvious that the origin of the M-I transition in the phase boundary (x≈0.3~0.4) is due to the phase separation and not to the structural phase transition. It is also found that the low-temperature orthorhombic phase in $Y_{0.63}Ca_{0.39}TiO_3$ can be identified as the metallic phase of this system.

The $Y_{1-X}Ca_XTiO_3$ (x=0.37, 0.39 and 0.41) samples used in this work were synthesized by the floating-zone method. The detail of the sample preparation will be described in elsewhere. The stoichiometry of the samples was carefully confirmed by EPMA. In order to characterize the electronic property of the samples, the temperature dependence of the electrical resistivity was measured by the four-terminal method for each sample from 1.5K up to 300K and from 300K down to 1.5K. Fig.1 shows the measured temperature dependence of the electrical resistivity. For the x=0.37 and 0.39 samples, the resistivity has the maximum around 60K and 130K, respectively, which indicate the M-I transition occurs at these temperatures. In addition, the thermal hysteresis is clearly observed indicating the transition is a first order type. On the other hand, the x=0.41 sample dose not show such a M-I transition at all and the monotonous behavior of the resistivity shows a metallic property from 1.5K to 300K. These temperature dependences of the resistivity are consistent with those of the previous studies.

For the precise crystal structure determination, the SR powder diffraction experiment was carried out by Large Debye-Scherrer Camera installed at BL02B2, SPring-8 [10]. The imaging plate was used as a detector to collect high counting statistics data. With this camera, both the high angular resolution and high counting statistics data can be collected. The powder samples for the X-ray experiment were prepared by crushing the high-quality single crystals. The granularity of the powder was made even less than 3-micron diameter by the precipitation method in order to get a homogeneous intensity distribution on Debye-Scherrer powder ring [11]. The obtained powder sample was sealed in a soda glass capillary (0.2mm in diameter). The He gas circulation type cryostat was used for the low temperature measurements. The X-ray powder patterns were measured from 300K down to 20K at intervals of 20K for each sample. All of the data were measured under the same experimental conditions except for the temperature. The exposure time of X-ray was 32 minutes for each temperature. The wavelength of incident X-rays was 0.735Å. The X-ray powder pattern was obtained with a 0.01°step from 8.0° to 74.0° in $2\theta$, which corresponds to 0.61Å resolution in *d*-spacing.

The obtained powder data were analyzed by the Rietveld method. In the present powder data, any impurity peaks were not found. The weighted profile reliability factors of the Rietveld refinement, $R_{WP}$, were in the range of 3 ~ 4% for all of the data. And the reliability factors based on the integrated intensities, $R_I$, were in the range of 3 ~ 5%. These good reliable factors mean that precise structure analyses are done satisfactorily for all the refinements.

For both x=0.37 and x=0.39, the observed powder pattern at 300K showed the orthorhombic *Pbnm* symmetry which has a distorted perovskite structure ($GdFeO_3$-type). On the other hand, the structure at 160K was the lower symmetry monoclinic $P2_1/n$. It is obvious that the orthorhombic to monoclinic structural phase transition has occurred at the temperature between 300K and 160K. In order to determine the orthorhombic-monoclinic transition temperature, the temperature dependence of the angle $\beta$ between a-c axes for x=0.37 and x=0.39 is plotted as shown in Fig.2. In this figure, the transition temperature can be determined at the temperature where the angle $\beta$ starts to deviate from 90 degrees. The plotted temperature dependence of $\beta$ values for the samples x=0.37 and 0.39 does not show significant difference. Then it is concluded that the orthorhombic (*Pbnm*) to monoclinic ($P2_1/n$) phase transition occurred at 230K for both x=0.37 and x=0.39 from Fig.2. The determined structural phase transition temperature is much higher than the M-I transition temperature 60K (x=0.37) and 130K (x=0.39), respectively. Any additional phase transition does not occur around the M-I transition temperature. Therefore, it is not possible to interpret the origin of the M-I transition by the phase transition found in the present study.

By the careful investigation of the powder data around the M-I transition temperature, we found the peculiar deformation of the intensity profile. For instance, the 020 profile of x=0.39 sample shows an anti-symmetric profile of a shoulder at the higher angle side around 140K, while the profile is symmetric at 300K. As lowering temperature, the 020 profile shoulder grows bigger than the original profile and finally became a main peak at 20K. The main peak at 300K shows an opposite tendency, i.e. the main peak turned to a lower angle shoulder at 20K. These facts indicate that two phases coexist below 140K and the volume ratio changes by lowering temperature. Thus, we have reanalyzed the low temperature data based on the mixed phases model. Fig.3 shows the fitting results of the Rietveld analysis of the data measured at 120K for x=0.39 sample. The main monoclinic phase and the low-temperature orthorhombic phase contributions are shown in Fig.3 (a) and (b), respectively. Over-all fitting results based on two-phase model are given in Fig.3 (c). It is clearly shown that the low-temperature orthorhombic phase has appeared below 140K. The mixed model gives the excellent fit with $R_I$=3.5% at 120K. Then, we have analyzed all of the measured powder patterns from 140K down to 20K based on the monoclinic and low-temperature orthorhombic two-phase model. The similar phase separation was also found for the sample x=0.37 below 60K. These phase separation temperatures show good agreement with their M-I transition temperatures.

The temperature dependence of the crystal structure of x=0.41 sample shows a very sharp contrast to those of x=0.37 and 0.39. The obtained powder data at all temperatures for this sample can be refined with the identical mixed phases model to x=0.39. This is consistent with the fact that there is no M-I transition for this sample and that the metallic behavior found for the sample x=0.39 at low temperature has preserved up to 300K. Consequently, there is no doubt that the M-I transition has closely related to the monoclinic + orthorhombic phase separation.

In the present study, the two kinds of orthorhombic phases are found for all the samples. In the case of x=0.39, one is above 230K and the other is below 140K. To see the role of the orthorhombic phases in the metallic (20K : low-temperature) and insulating (300K : high-temperature) phases, we have examined the difference of the lattice parameters between the low-temperature and high-temperature orthorhombic structures. Fig.4 shows the phase diagram and the temperature dependence of the lattice parameters for x=0.39 determined in this study. It is found that the lattice parameter, b, of the low-temperature orthorhombic phase is distinctly shorter than that of the high-temperature orthorhombic phase. By taking thermal expansion into account, the other two lattice parameters, a and c, are virtually the same with those of the high-temperature phase. The similar difference of the lattice parameters between the orthorhombic phases was also observed for the x=0.37 sample. The lattice parameters of the low-temperature orthorhombic phase of the sample x=0.41 coincide with those of the low-temperature orthorhombic phases in x=0.37 and x=0.39 samples. The lattice parameters of each orthorhombic phase at 20K and 300K are listed in Table 1 with those of the monoclinic phase at 20K for each sample. As for the monoclinic phase, there is no lattice parameter jump relating to the M-I transition. These facts indicate that the low-temperature orthorhombic phase is not identical to that of the high-temperature orthorhombic phase and should play a crucial role for the metallic property at low temperature.

To see the contribution of the separated phases to the M-I transition in detail, the volume ratio of the monoclinic phase are shown with the temperature dependence of the electrical resistivity on each sample in Fig.5. The values of the volume ratio were determined by the Rietveld refinement. By comparing the volume ratio of the monoclinic phase with the temperature dependence of the resistivity, it was found that the decrease of the volume ratio had a strong correlation with the temperature dependence of the resistivity. From the facts described above, it can be concluded that the low-temperature orthorhombic phase contributes to the metallic property of this system. In other words, the M-I transition seems to occur as a result of the appearance of a certain amount of the metallic low-temperature orthorhombic phase in the insulating monoclinic phase.

In summary, the following structural aspects of this system were found by the precise SR powder structure analyses.

1. For both $Y_{0.63}Ca_{0.37}TiO_3$ and $Y_{0.61}Ca_{0.39}TiO_3$, the orthorhombic (*Pbnm*) to monoclinic (*P2$_1$/n*) phase transition occurs at around 230K, which is much higher than the M-I transition temperature.
2. The significant phase separation occurs in the vicinity of the M-I transition temperature for both $Y_{0.63}Ca_{0.37}TiO_3$ and $Y_{0.61}Ca_{0.39}TiO_3$.
3. As for $Y_{0.59}Ca_{0.41}TiO_3$, the structure has already separated into two phases at 300K.

In conclusion, we found the phase separation in the phase boundary region ($x \approx 0.3 \sim 0.4$), which cause the M-I transition. In the mixed phases at low-temperature region, the orthorhombic phase must be metallic and the monoclinic phase must be insulating. This assignment explains the experimental observations of the electrical resistivity of this system. We also found the structural phase transition at much higher temperature than the M-I transition temperature. This phase transition does not contribute to the M-I transition.

The similar phase separation has been also observed in perovskite-type manganites. In the manganites system, there have been controversies whether the phase separation is the origin of both the M-I transition and the colossal magnetoresistance (CMR). For example, the doped manganites, $Nd_{0.55}(Sr_{0.17}Ca_{0.83})_{0.45}MnO_3$, has the phase separated state below the charge ordering temperature [12]. On the other hand, in the perovskite, $YNiO_3$, it has been reported that the orthorhombic (*Pbnm*) to monoclinic (*P2$_1$/n*) phase transition has simultaneously occurred with the M-I transition [13]. The metallic and insulating phases in this system are identified as the orthorhombic and monoclinic phases, respectively. Such a complicated structural varieties make it difficult to identify the origin of the M-I transition for these materials. The present work suggests the importance of the precise structural studies for both metallic and insulating phases of these materials to understand the origin of the M-I transitions.

Acknowledgements
This work has been partially supported by a Grant-in-Aid for Science Research from the Ministry of Education, Science, Sports and Culture, Japan. This work was also supported by the Nippon Sheet

Table 1 The lattice parameters for $Y_{1-x}Ca_xTiO_3$ (x=0.37, x=0.39 and x=0.41) at 20K and 300K

| Composition x | | 0.37 | | 0.39 | | 0.41 | |
|---|---|---|---|---|---|---|---|
| Temperature (K) | | 20 | 300 | 20 | 300 | 20 | 300 |
| High-temperature orthorhombic to monoclinic phase (Insulating) | Space Group | *P2₁/n* | *Pbnm* | *P2₁/n* | *Pbnm* | *P2₁/n* | *Pbnm* |
| | a (Å) | 5.35001(5) | 5.35621(3) | 5.3495(2) | 5.35622(2) | 5.3523(6) | 5.3598(2) |
| | b (Å) | 5.58185(5) | 5.57740(3) | 5.5772(2) | 5.57210(2) | 5.5707(5) | 5.5702(2) |
| | c (Å) | 7.64427(8) | 7.65815(4) | 7.6414(3) | 7.65838(3) | 7.6451(8) | 7.6608(2) |
| Low-temperature orthorhombic phase (Metallic) | Space Group | *Pbnm* | | *Pbnm* | | *Pbnm* | *Pbnm* |
| | a (Å) | 5.3445(1) | | 5.34059(3) | | 5.34149(4) | 5.35608(5) |
| | b (Å) | 5.5551(1) | | 5.55120(3) | | 5.54651(4) | 5.56298(5) |
| | c (Å) | 7.6486(2) | | 7.64535(6) | | 7.64652(6) | 7.65909(8) |

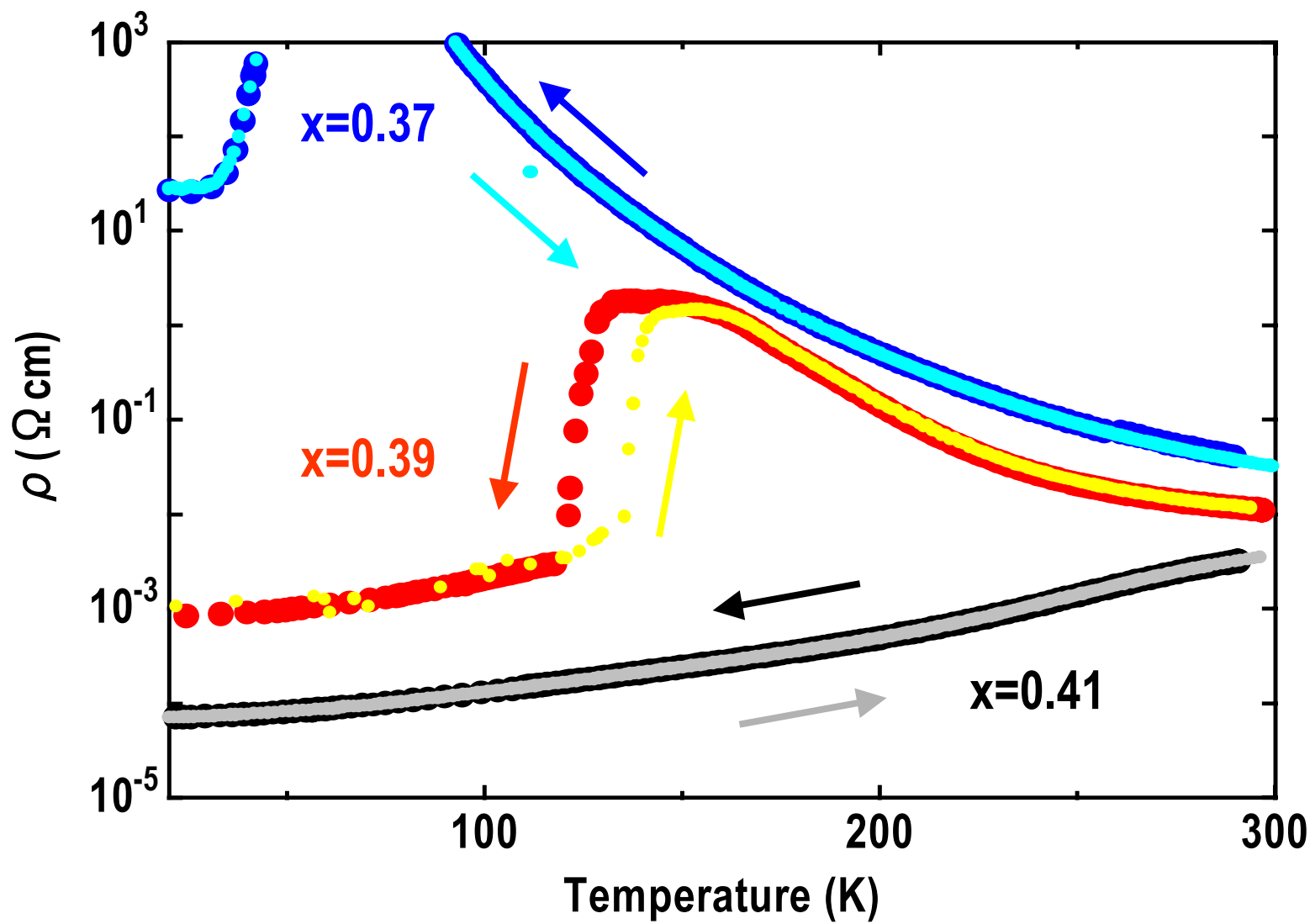

**Fig.1** Temperature dependence of the electrical resistivity for $Y_{1-x}Ca_xTiO_3$ (x=0.37, 0.39 and 0.41)

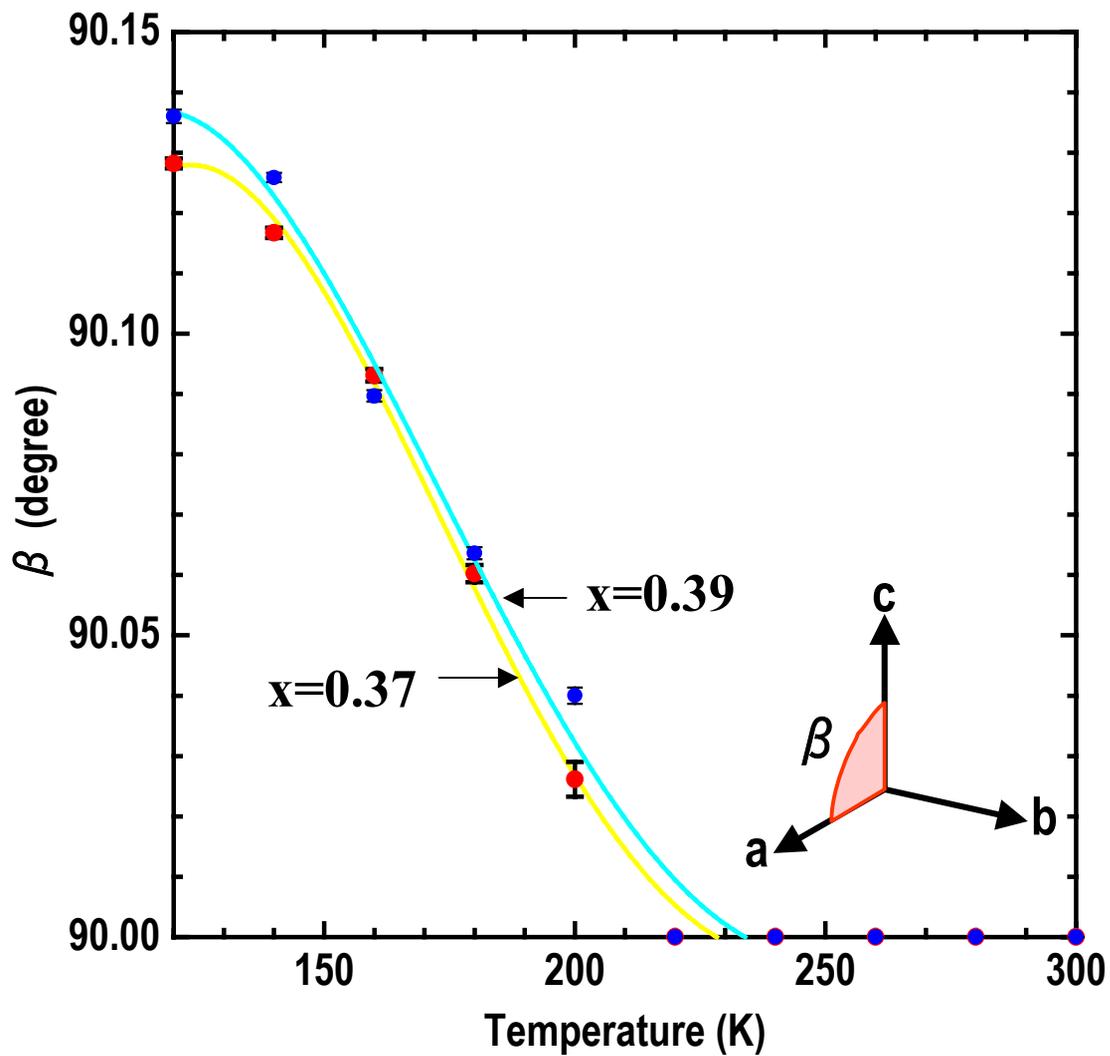

**Fig.2 Temperature dependence of the monoclinic angle $\beta$ for x=0.37 and 0.39**

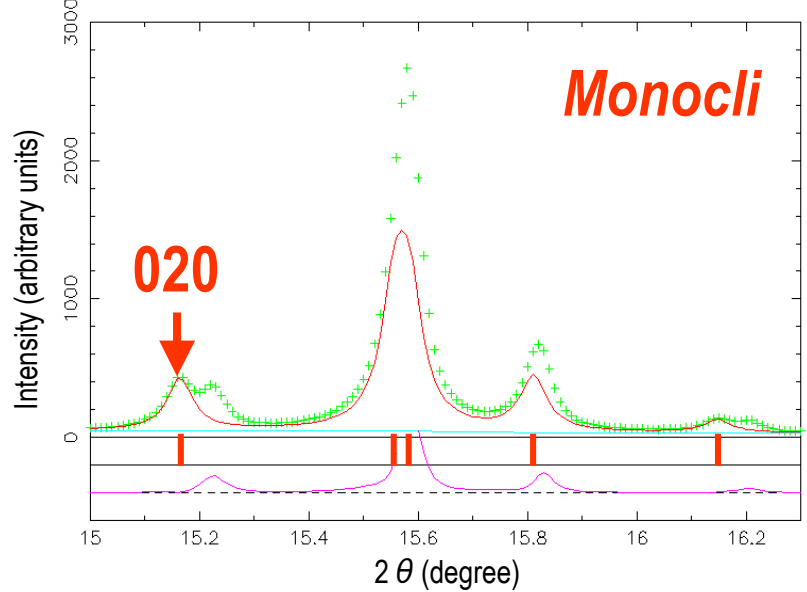 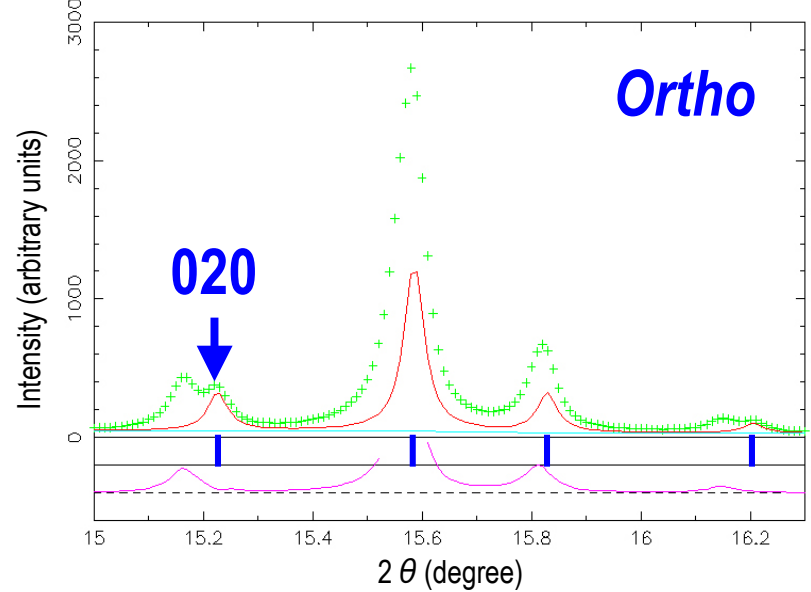

**(a)** **(b)**

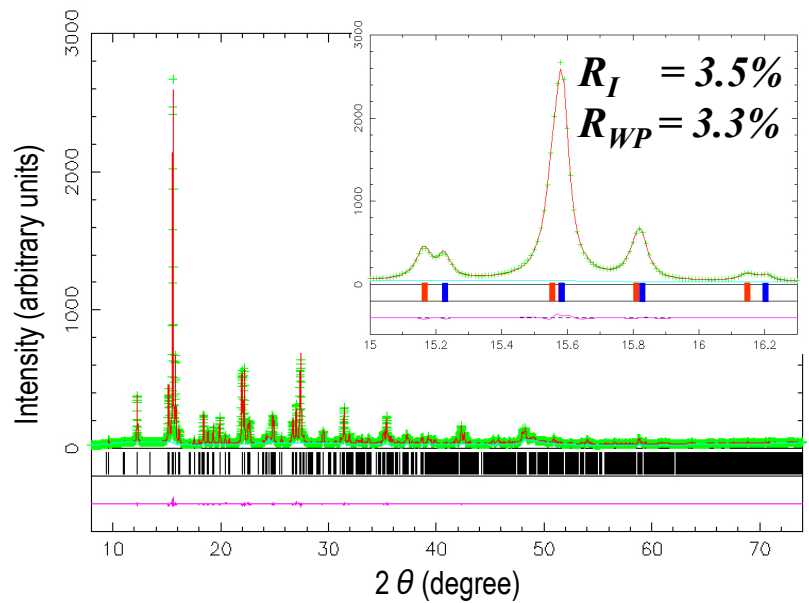

**(c)**

**Fig.3 The fitting results of Rietveld analysis for x=0.39 at 120K**

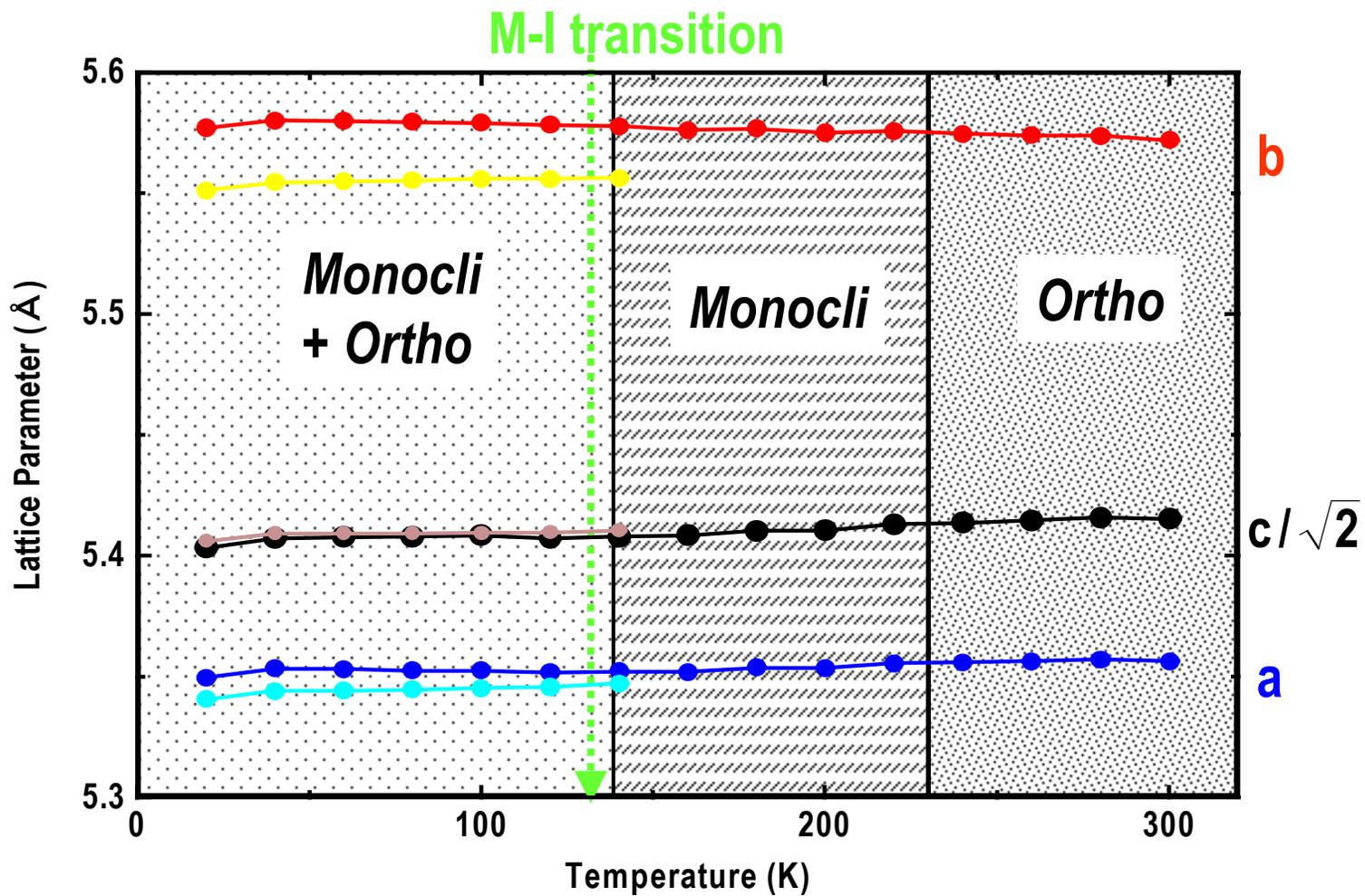

Fig.4 Phase diagram with temperature dependence of lattice parameters for x=0.39

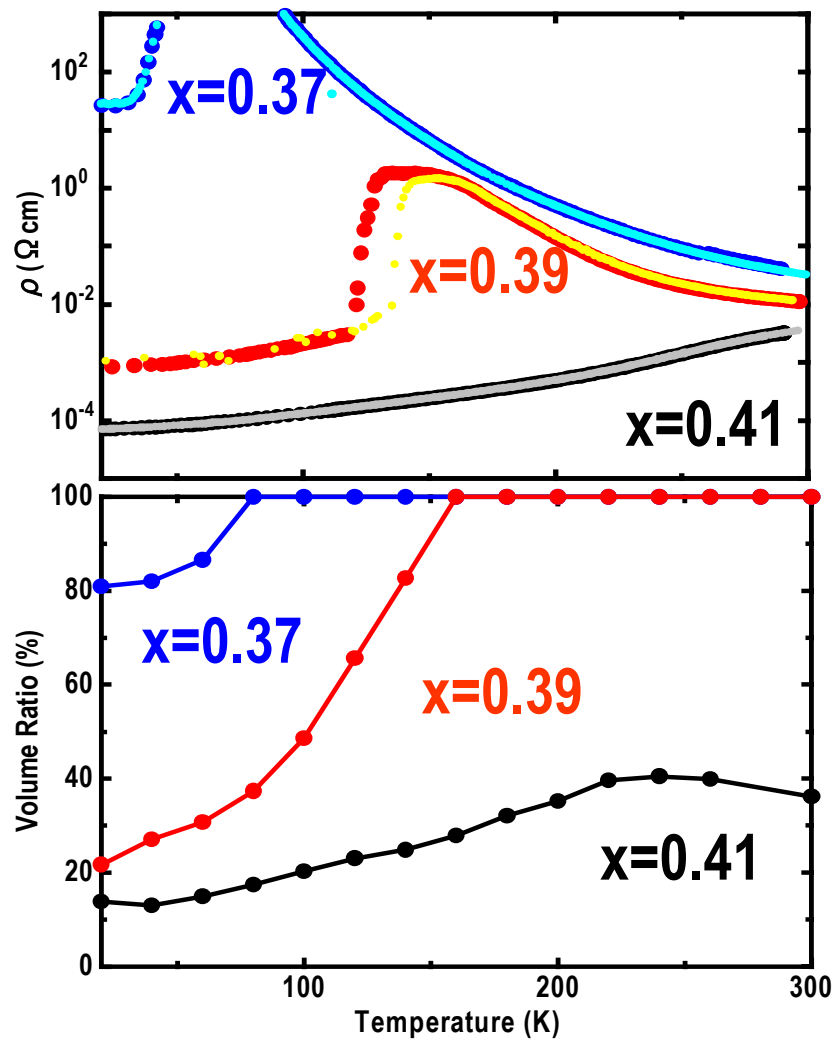

Fig.5 Temperature dependence of the volume ratio of the monoclinic phase for x=0.37, 0.39 and 0.41 compared with temperature dependence of the electrical resistivity